\documentclass[11pt]{article}

\usepackage[margin=1in]{geometry}

\usepackage{fontspec}
\newfontfamily\arabicfont[Script=Arabic]{FreeSerif.ttf}

\usepackage{graphicx}
\usepackage{longtable}
\usepackage{array}
\usepackage{tabularx}
\usepackage{ragged2e}
\usepackage{booktabs}
\usepackage{caption}
\usepackage{enumitem}
\usepackage[hidelinks]{hyperref}
\usepackage{xurl}
\usepackage{footmisc}
\usepackage{titlesec}
\usepackage{fancyhdr}
\usepackage{abstract}
\usepackage{xcolor}
\usepackage{bidi}

\newcommand{\ar}[1]{%
  \parbox[t]{\linewidth}{%
    \setRTL
    \arabicfont
    #1%
  }%
}
\hypersetup{
  colorlinks=true,
  linkcolor={red!55!black},
  citecolor={red!55!black},
  urlcolor=blue,
  pdftitle={Perceived and Actual AI Deepfakes: The Case of Sudan},
  pdfauthor={Eilaf Mohamed}
}

\newcolumntype{Y}{>{\RaggedRight\arraybackslash}X}
\newcolumntype{L}[1]{>{\RaggedRight\arraybackslash}p{#1}}

\newcommand{\refentry}[2]{\hypertarget{cite:#1}{}\par\noindent\hangindent=1.5em\hangafter=1 #2\par\smallskip}

\title{\textbf{Perceived and Actual AI Deepfakes: The Case of Sudan}}
\author{Eilaf Mohamed\thanks{Published as part of the 2025 AMEL Sudan Democracy Lifeline Fellowship under the supervision of Hamid Khalafallah. }}
\date{}

\begin{document}

\maketitle
\thispagestyle{fancy}

\begin{abstract}
Sudan's AI deepfake risk currently appears driven more by demand-side vulnerabilities than the volume of AI deepfake content. While AI-generated disinformation in Sudanese feeds remains limited\footnote{This statement holds as of the time of writing this paper.}, perceived (alleged) AI deepfakes and general AI skepticism worsen the situation and contribute to general uncertainty in multimedia content. Within the analyzed cases, we found that the prominent stance was distrust driven more by motivated reasoning and contextual reliance than by durable AI detection skills. In practice, rejection or acceptance of AI deepfakes often reflects belief-driven judgment more than technical assessment\footnote{These findings should be interpreted with caution given the limitations of inferring beliefs from social media comments. Refer to the limitations section.}. These weak defenses leave the information environment vulnerable to future sophisticated AI deepfake campaigns and risk normalizing truth indifference. Short-term interventions should prioritize modality-agnostic demand-side interventions such as pre-bunking, AI literacy, scaled fact-checks, and tiplines that strengthen public resilience across all forms of disinformation. More technical AI-specific and supply-side interventions can be developed as institutional capacity and technology allow.
\end{abstract}

\section{Introduction}

Deepfake is the process of creating fake digital content or using deep learning techniques to manipulate authentic materials; the content consists of text, audio, images, or video (\hyperlink{cite:gambin2024}{Gambín et al., 2024}). A central concern in the literature on the societal impacts of deepfakes is the use of synthetic media in disinformation campaigns, as visual and auditory communication establishes fluency due to the sense of familiarity it gives to the viewers; therefore, the content is perceived to be more truthful (\hyperlink{cite:helmus2022}{Helmus, 2022a}; \hyperlink{cite:vaccari2020}{Vaccari \& Chadwick, 2020}). AI deepfakes pose security and political risks because abundant public footage of politicians lets malicious actors easily create high-quality deepfakes for manipulating elections, escalating ethnic divides (\hyperlink{cite:helmus2022}{Helmus, 2022a}; \hyperlink{cite:schwartz2018}{Schwartz, 2018}; \hyperlink{cite:vaccari2020}{Vaccari \& Chadwick, 2020}), and reducing trust in institutions (\hyperlink{cite:helmus2022}{Helmus, 2022a}). In conflict settings, AI deepfakes can be used to legitimize war and uprisings, falsify orders, undermine popular support, polarize societies, divide allies, and discredit leaders (\hyperlink{cite:byman2023}{Byman et al., 2023}).

One of the societal impacts of AI deepfakes is increasing uncertainty and undermining trustworthy sources of information (\hyperlink{cite:helmus2022}{Helmus, 2022a}; \hyperlink{cite:vaccari2020}{Vaccari \& Chadwick, 2020}). This confirms \hyperlink{cite:schwartz2018}{Schwartz's (2018)} theory that ``the greatest threat is not that people will be deceived, but that they will come to regard everything as deception,'' which some scholars call an epistemological threat. Their theory builds on Downs' (1957) theory that uncertainty is created because of the high cost of accessing information. Deepfakes increase the cost, as users become unable to verify information (\hyperlink{cite:vaccari2020}{Vaccari \& Chadwick, 2020}). Extending (\hyperlink{cite:liu2025}{Liu et al., 2025}) framing of demand side vulnerabilities, in this paper we argue that truth becomes indifferent, not just as a passive response to misinformation, but also because the users are biased in their consumption of information.

Adding to the complexity of the problem, perceived or alleged AI deepfakes go beyond technical manipulation; while deepfakes promote belief in a falsified truth, perceived deepfakes cultivate a general AI deepfake skepticism to reject genuine truth. In that sense, deepfakes can impact objective truth when people refuse unfavorable correct information by assuming that it might be a deepfake, based on what scholars refer to as the liar's dividend (\hyperlink{cite:helmus2022}{Helmus, 2022a}; \hyperlink{cite:schiff2024}{Schiff et al., 2024}). In this paper we explore a different variant of this problem, one that is led by demand side vulnerabilities, where the recipients of information reject truthful material and perceive it to be deepfake. We use the term ``Perceived AI deepfake'' to refer to this phenomenon.

The models of disinformation help understand the impact of AI deepfakes and develop targeted interventions. The passive aggregator view, ``Disinformed by others,'' suggests that the public passively absorb disinformation from malicious actors, which requires interventions to focus on the supply side, whereas the motivated public view, ``Misinformed by themselves,'' calls for demand-side interventions and argues that, instead of disinformation shaping beliefs, disinformation is shaped by motivated reasoning (\hyperlink{cite:kahan2017}{Kahan, 2017b}), cognitive abilities (\hyperlink{cite:pennycook2018}{Pennycook \& Rand, 2018}; \hyperlink{cite:ahmed2021}{Ahmed, 2021}), AI self-efficacy (\hyperlink{cite:liu2025}{Liu et al., 2025}), and other individual-level factors that increase the acceptability of disinformation. The cognitive abilities account found that people were more likely to believe in fake news\footnote{The study was conducted on textual fake news.} because of low Cognitive Reflection Test (CRT) scores, regardless of political ideology (\hyperlink{cite:pennycook2018}{Pennycook and Rand, 2018b}), whereas the motivated reasoning shows that partisan or political identity influences cognition, memory, genuine evaluation, and perceptions (\hyperlink{cite:vanbavel2018}{Van Bavel \& Pereira, 2018}). \hyperlink{cite:liu2025}{Liu et al.\ (2025)} introduce a new dimension to reality apathy, or truth indifference, which is AI self-efficacy, and they found that low AI self-efficacy resulted in higher cynicism. While all three dimensions of demand side vulnerabilities offer valuable insights, this paper aligns more closely with the motivated reasoning perspective\footnote{Because it was straightforward to observe within the context of this study.} as it offers better explanation to the Sudanese public reaction to AI deepfake.

Despite global concern over AI deepfakes and AI's societal impact in general, limited research investigates the public reaction to adversarial AI deepfakes in the context of fragile states, a high-stakes context with limited trust in institutions, high information gate-keeping, limited AI literacy, and escalated virtual propaganda. This study aims to understand the impact of AI deep fakes on in Sudan through empirical research and also explore feasible mitigation strategies by answering the following research questions:

\begin{itemize}[leftmargin=2em]
\item RQ1: What is the public reaction to perceived and actual AI deepfakes used in Sudan's conflict
\item RQ2: What factors drive the truth assessment of perceived and actual AI deepfakes?
\item RQ3: What are the possible interventions to address AI deepfake in the Sudanese information environment?
\end{itemize}

The paper starts by explaining the problem of AI deepfake in the Sudanese media, we analyze two cases that included either deepfakes or perceived deepfakes from March 2023 until March 2024, followed by a case study on Hmedti's death misinformation campaign. We then use the insights from our results to reframe the problem and establish a theoretical framework for the role of motivated reasoning in the AI deepfake problem. Finally, we use the insights from our personal interviews and the disinformation literature to offer recommendations for addressing AI deepfakes in Sudan.

\section{AI Deepfake Problem in Sudan}

In April 2023, the conflict has started in Sudan between the government of Sudan, controlled by the Sudanese Armed Forces (SAF) under General Abdel Fattah al-Burhan against the Rapid Support Forces (RSF) paramilitary group, led by General Muhammad Hamdan Dagalo, also known as Hemedti. The UN has reported that the conflict led to one of the worst displacement and famine crises in the world. During the conflict, the online space in Sudan has been fueled by misinformation campaigns to shape the narratives that each rival force is trying to impose on the eyes of the public with Rapid Support Forces portrays itself as a protector of the revolution, fighting Islamic radicals, and Sudanese Armed Forces (SAF) use the notion of defends Sudan's sovereignty and rebels. Islamists also use online spaces to call for war continuation and incite civilian leaders against SAF (\hyperlink{cite:khalafallah2025}{Khalafallah, 2025}). The deepfakes have already been used to support the narratives of conflicting parties and mislead public opinion. While most of the disinformation supply in Sudan was either textual fake news or out-of-context disinformation, there were some instances where AI was used to escalate virtual propaganda through voice cloning and AI deepfake technologies. The earliest one was in October 2023, when a TikTok account posted fake voice cloning recordings for Omer Al-Bashir---the ex-president---criticizing current military leaders (\hyperlink{cite:goodman2023}{Goodman \& Hashim, 2023})\footnote{Following the BBC investigation, TikTok removed the account.}. Another attempt followed in March 2024, when a voice cloning featuring Commander Abdel Fattah al-Burhan and Abdeen al-Shami discussing plans to kill civilians circulated through the X platform (\hyperlink{cite:beam2025a}{Beam Reports, 2025a}).

Notably, the public also participated in AI deepfake generation, either for sarcasm or to raise awareness. A March 2024 public-created recording featuring the Rapid Support Commander (RSF), Hemedti, with leaders of the Forces for Freedom and Change (FFC), revealed their conspiracy before April 2023 and circulated through the X platform (\hyperlink{cite:beam2025a}{Beam Reports, 2025a}). Although the footage was declared to be AI-generated, it was later shared as an authentic recording, and the National TV also contributed to promoting this propaganda by sharing the AI-generated media (\hyperlink{cite:beam2025a}{Beam Reports, 2025a}; [1]\footnote{Bracketed numbers in the text refer to personal interviews, which are detailed in Annex A.}). The normalization of synthetic media in official communication is exemplified by the AI-generated video of Jebel Marra that has been circulated through the Sovereign Council's official social media platforms [1]. 

In Sudan, AI deepfake skepticism, or perceived AI deepfakes, is a prominent problem. It occurs when footage is dismissed as AI deepfakes, either by politicians to support their agendas (liar dividend) or by the public to support their biases (perceived deepfake). The manipulation of belief by dismissing real content as a deepfake can be illustrated by the classic instance of the liar dividend in June 2023, when a recording of a conversation between Mubark Ardol and other political figures was leaked. They responded by claiming that it was an AI deepfake and generated by their rivalry (\hyperlink{cite:suliman2024}{Suliman, 2024}).

More generally, the information environment in Sudan makes a perfect incubator for AI deepfakes with limited trust in media outlets and weak legal infrastructure. The information environment has been seen by successive authoritarian regimes as a dangerous tool that enables mobilization for protesters, and therefore, the legal infrastructure is weak and designed to suppress freedom of speech and enable harmful acts against the opposition (\hyperlink{cite:knight2023}{Knight \& Alsedeg, 2023}). Although there exists legal infrastructure, such as the Access to Information 2015 law, which grants citizens the right to access information, but it has vague language, and activists argued that it protects the interests of those in power, which explains the effect of information poverty (\hyperlink{cite:knight2023}{Knight \& Alsedeg, 2023}, [2]).

The impact of AI deepfake in the information environment has a sensitive context in fragile states as they can go beyond persuasion, including decreased level of accountability and scrutiny when it comes to publishing news. The public will be less engaged in fact-checking because they believe everything on social media is false. As a result, it will be hard to have active and informed discussions on public affairs. This can threaten any potential form of democracy, as it requires citizens to be informed and actively engaged. Another impact is the potential use of the deepfakery environment by authoritarian regime to suppress freedom of expression, a scenario where the political elite might campaign for more restrictive media outlets that do not respect freedom of speech, assuming that it will help restore order (\hyperlink{cite:schiff2024}{Schiff et al., 2024}; \hyperlink{cite:vaccari2020}{Vaccari \& Chadwick, 2020}).

\section{Methodology}

We attempted to answer RQ1 and RQ2 through a between-subject trust/uncertainty measurement for X (formerly Twitter) comments, comparing public reaction to four tweets accompanied by AI deepfake footage. The tweets\footnote{Annex B provide a detailed citation and documentation of the referenced tweets.} were posted from April 2023 until April 2024, comparing:

\begin{enumerate}[leftmargin=2em]
\item Confirmed deepfake: Jonny's tweet, Salah Manaa's 2nd Tweet, Obai and Yassir Alata's tweet
\item Perceived deepfake: Ardol's tweet, Hemedti's death tweets
\item Authentic: Salah Manaa's 1st Tweet
\end{enumerate}

The top 80--100 text-based comments per tweet were extracted via Apify, and the data was ranked based on the ``most liked'' comments to create a representative sample. The data cleaning excluded  comments containing prophane languge\footnote{While profanity may reveal insights about public sentiment we found their usefulness to be limited}. To ensure the reliability of the coding, we adopted three labeling passes to ensure stabilized codes and documented the rules and assumptions underlying the labeling process. Obai and Yassir Alata's tweet were later removed from the sample because they shared footage that was already labelled as deepfake and therefore they revealed limited insights on the public reaction to deepfake material when they are exposed to it for the first time. Similarly, Hmedti's death tweets were later removed from the sample because it was hard to trace the first tweet on the deepfakery of his footage and the sentiment around this claim has shifted overtime. Instead, a case study approach based on desk review of secondary sources was adopted to study the public reaction to Hmedti's death disinformation. The labelling process involved two stages:

\subsection*{1. Labeling trust/uncertainty}

Given the novelty and controversial nature of comments in political contexts, we found that reactions may not be limited to the perceived authenticity of the footage but also extend to the context or narrative presented within the tweet. Therefore, the trust labeling was holistic. Table 1 show the meaning of each label.

\begin{longtable}{@{}>{\RaggedRight}p{0.22\textwidth}@{\hspace{1.5em}}>{\RaggedRight\arraybackslash}p{0.68\textwidth}@{}}
\caption{Trust/uncertainty labels}\label{tab:trust-labels}\\
\toprule
\textbf{Label} & \textbf{Meaning} \\
\midrule
\endfirsthead
\multicolumn{2}{@{}l}{\textit{Table \thetable{} continued}}\\
\toprule
\textbf{Label} & \textbf{Meaning} \\
\midrule
\endhead
\bottomrule
\endfoot
Distrusting & ``Distrusting'' label was given to all comments expressing distrust, whether of the footage or the overall narrative. \\
Trusting & The ``trusting'' label was given to comments that expressed trust in both the media and the narrative or explicitly trusted the narrative without questioning the authenticity of the footage. \\
Irrelevant & The ``irrelevant'' label was given to the comments that revealed limited insights into the user's perception of the truthfulness of the footage and the overall tweet. \\
Indifferent & The ``indifferent'' label was given to those who perceived the truth or deepfakery of the footage to be irrelevant. \\
\end{longtable}

\subsection*{2. Multi-thematic labelling For comments with distrusting label}

To answer RQ2 we adopted multi-label theme analysis\footnote{The labels were grounded on the insights made from the first pass of the comments for each tweet.} to better understand the comments with the majority label (Distrusting label). The themes vary across tweets, with some overlapping themes. Each comment was given between one and three themes, given the embedded meaning of the comment.

\begin{longtable}{@{}>{\RaggedRight}p{0.22\textwidth}@{\hspace{1.5em}}>{\RaggedRight\arraybackslash}p{0.68\textwidth}@{}}
\caption{Qualitative Themes in distrusting comments}\label{tab:themes-distrust}\\
\toprule
\textbf{Tweet} & \textbf{Thematic labels for distrusting comments} \\
\midrule
\endfirsthead
\multicolumn{2}{@{}l}{\textit{Table \thetable{} continued}}\\
\toprule
\textbf{Tweet} & \textbf{Thematic labels for distrusting comments} \\
\midrule
\endhead
\bottomrule
\endfoot
Jonny's tweet & Agree with the media but disagree with the context/narrative; AI usage; Critical analysis/fact inconsistency; other; Source distrust; Support for the subject of the video \\
Ardol's tweet & disbelieve AI usage; Call for Accountability; Source distrust; Other; Critical analysis/fact inconsistency; Unlikely Target (Low Prominence); Mocking Official Language \\
\end{longtable}

For answering RQ3 we performed a disk review and conducted semi structured interviews with civic tech practitioners and fact organizations in Sudan. More details about the interviewees will be found in Annex B.

\section{Results}

\subsection{Actual Deepfake: Jonny's tweet}

On March 15th, 2024, Jonny Gould, an Israeli news presenter, shared a tweet with a voice recording of what he alleged to be a conversation between al-Burhan and his Deputy Chief of Operations. He accused Gen.\ al-Burhan of orchestrating an attack on civilians and ordering soldiers to occupy their homes. The caption described Gen.\ al-Burhan as the head of the Muslim Brotherhood in Sudan. He ended the caption by criticizing the global response, attributing their inaction to Israel's non-involvement. Investigations by Beam Mersad found the recording to be AI-generated (\hyperlink{cite:beam2025a}{Beam Reports, 2025a}).

Our analysis of the most liked comments (n=106) revealed the following:

While the majority (82.1\%) distrusted the tweet, 9.4\% of the commenters trusted it, indicating limited but not negligible deception.

\begin{figure}[htbp]
\centering
\includegraphics[width=0.85\linewidth]{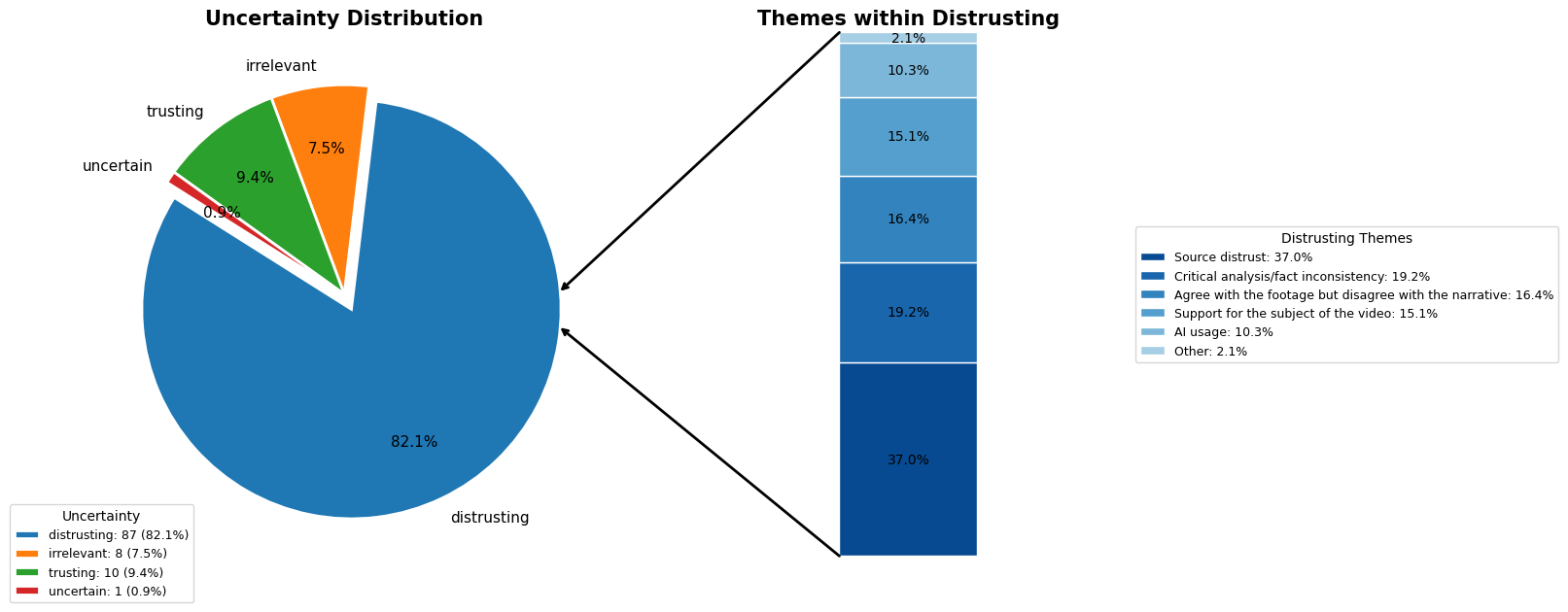}
\caption{Uncertainty Distribution of the Comments on Jonny's Tweet}
\end{figure}

Table 3 below presents the recurring themes\footnote{For more details on the coding framework, refer to Annex C.}, particularly for those who cited distrust. We found that the majority of the themes (37\%) were related to the source's reputation and speculating on conspiracy and corrupted intent. This was followed by a critical analysis of the tweet narrative (19.2\%). The most interesting primary theme was ``Agree with the footage but disagree with the context/narrative,'' a theme in which commenters perceived the media to be truthful but attempted to contextualize it by responding to the caption. This was frequently linked to their analysis of facts or at least their perception of it, which in many cases correlated with their support for the subject of the recording. The AI usage theme appeared in only 10.3\% of the comments, highlighting that truth judgment was largely influenced by source cues and motivated reasoning rather than the detection skills of the technical deepfakery of the footage. Table 3 shows the themes, their distribution, and representative quotes from the dataset.

\begin{longtable}{@{}>{\RaggedRight}p{0.32\textwidth}@{\hspace{1em}}>{\RaggedRight}p{0.12\textwidth}@{\hspace{1em}}>{\RaggedRight\arraybackslash}p{0.43\textwidth}@{}}
\caption{Qualitative Themes in the Comments on Jonny's Tweet}\label{tab:jonny-themes}\\
\toprule
\textbf{Theme} & \textbf{Percentage} & \textbf{Quote} \\
\midrule
\endfirsthead
\multicolumn{3}{@{}l}{\textit{Table \thetable{} continued}}\\
\toprule
\textbf{Theme} & \textbf{Percentage} & \textbf{Quote} \\
\midrule
\endhead
\bottomrule
\endfoot
Source distrust & 37\% & ``You and your fellow Jewish media friends support the crimes of the RSF'' \\
Critical analysis/fact inconsistency & 19.2\% & ``Abdul Fattah Al Burhan isn't the head of Muslim brotherhood'' \\
Agree with the footage but disagree with the narrative & 16.4\% & ``All the buildings alburhan said located around military sites and definitely RSF militia will bomb them first, so it's better to be used by army.'' \\
Support for the subject of the recording & 15.1\% & ``We fully support our Sudanese military'' \\
AI usage & 10.3\% & ``It's obviously fake'' \\
other & 2.1\% & \\
\end{longtable}

The results highlight that people generally distrusted the AI deepfake tweets but mostly due to the misalignment between the tweet narrative and their political views. In some cases people believed the AI deepfake but they rejected the narrative in which it was shared and attempted to contextualise it, in line with their political views. This is an important theme of the interaction of AI deepfake footage with the demand side. The results highlight the gap in the public detection skills of AI generated content and the strong influence of their biases in their truth judgement of AI deepfake content.

\subsection{Perceived AI deepfake: Ardol's tweet}

On July 16, 2023, Mubarak Ardol, a Sudanese politician, posted what he called a ``Joint Public Statement'' between him and Nour Al-Daem Taha. He posted this statement in response to a circulated recording on social media of them with Abdelwahab Jameel, in which they shared problematic views on their position toward the 2018 revolution and notable political leaders. In his tweet, he stated that the footage was AI-generated. He argued that the public accessibility of their voice samples and the advancement of artificial intelligence technologies enabled the oppositional rallying to produce this AI deepfake footage. He also made an implicit inference to Hmedti's death, saying, ``It is now easy to imitate anyone's voice, even those who are deceased,'' as evidence for the usage of AI deepfakes in manipulating public opinion. Evidence from \hyperlink{cite:suliman2024}{Suliman (2024)} and the public investigations on the footage proved that the recording was authentic.

Figure 2 (n=105) shows that while 91.4\% distrusted the tweet, 2.9\% cited uncertainty, which is larger than the cited uncertainty in Jonny's tweet. The smallest proportion (1.9\%) trusted the tweet and believed in AI usage, lower than in Jonny's tweet. This implies that those who were deceived by the perceived/alleged deepfake were relatively fewer than those deceived by actual deepfakes.

The qualitative analysis showed that the majority rejected AI usage, believing the recording was too sophisticated to be AI-generated, with insider jokes, local jargon, and voice inference, which they believed was beyond AI generative capabilities at the time, possibly reflecting some level of AI literacy skills\footnote{Comments may not accurately reflect AI literacy. Refer to the limitations section.}. Recurring themes included ``Call for accountability'' and ``Source distrust,'' in which people referred to his negative reputation to make judgments about the truthfulness of his claims. Additional themes included speculating on fact consistency or speculating on his prominence. These themes suggest that the narrative support and public controversy of the actor may influence the truth judgement of the content.

\begin{figure}[htbp]
\centering
\includegraphics[width=0.85\linewidth]{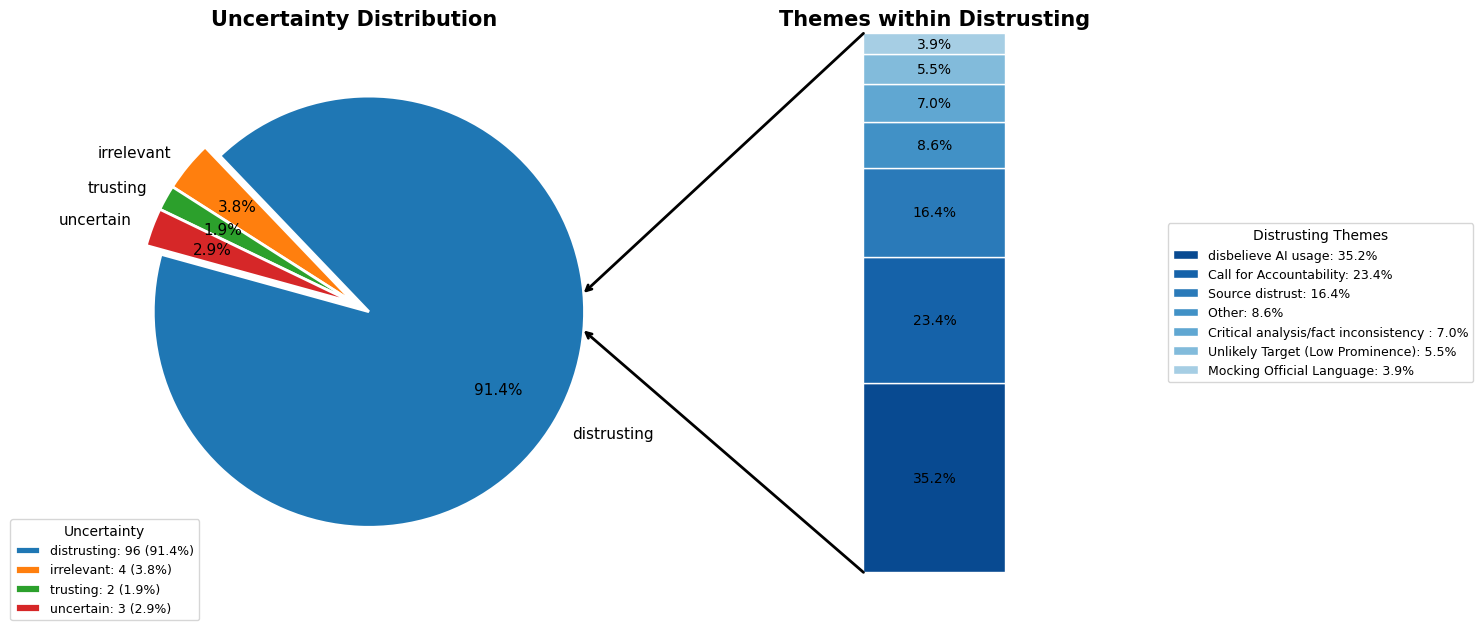}
\caption{Uncertainty Distribution of the Comments on Ardol's Tweet}
\end{figure}

\begin{longtable}{@{}>{\RaggedRight}p{0.15\textwidth}@{\hspace{0.8em}}>{\RaggedRight}p{0.10\textwidth}@{\hspace{0.8em}}>{\RaggedRight}p{0.29\textwidth}@{\hspace{0.8em}}>{\RaggedRight\arraybackslash}p{0.32\textwidth}@{}}
\caption{Qualitative Themes in the Comments on Ardol's Tweet}\label{tab:ardol-themes}\\
\toprule
\textbf{Theme} & \textbf{Percent.} & \textbf{Quote} & \textbf{Translated Quote} \\
\midrule
\endfirsthead
\multicolumn{4}{@{}l}{\textit{Table \thetable{} continued}}\\
\toprule
\textbf{Theme} & \textbf{Percent.} & \textbf{Quote} & \textbf{Translated Quote} \\
\midrule
\endhead
\bottomrule
\endfoot
disbelieve AI usage & 35.2\% &
\ar{عشان تعمل حوار بين ثلاثة أنفار ( وفي الخلفيه جي بي أس) الثانيه كم واربعين وبرضوا بعد دقيقه وكم خمسين يوضح صوتين متداخلين بصورة يصعب علي الذكاء الاصطناعي في هذه المرحلة القيام بها}
&
creating a conversation between three individuals (with GPS in the background) that lasts for about forty-something seconds, and then---after a minute and fifty-something---two overlapping voices appear in a way that AI, at its current stage, finds difficult to replicate. \\[4pt]
Call for Accountability & 23.4\% &
\ar{لمن تقول كلام ابقى قدره}
&
If you say something, be able to stand by it. \\[4pt]
Source distrust & 16.4\% &
\ar{البيك معروفه من زمان يعني مابقت على التسجيل}
&
Your true nature has been known for a long time, so it's not just about the recording \\[4pt]
Other & 8.6\% & & \\[4pt]
Critical analysis/fact inconsistency & 7.0\% &
\ar{نفس كلامك ده قلتو وقت ده هتف و اتهمته قحت وره الكلام و بعد الحرب قلت قحت مع الجنجويد و اسي نفس الشي بتعملو}
&
You said the same thing back then when you cheered and accused the FFC (Forces of Freedom and Change) of being behind it. Then after the war, you said the FFC was with the Janjaweed---and now you're doing the exact same thing again. \\[4pt]
Unlikely Target (Low Prominence) & 5.5\% &
\ar{لو كنت زول عندك مكانة والا سمعة والا وجودك مؤثر كان قلنا دا سبب إنو صوتك يتفبرك وبي ذكاء اصطناعي لكن أنت ولا شي}
&
If you were someone with status, a reputation, or any real influence, we might've said that's why your voice was faked with AI. But you're nothing. \\[4pt]
Mocking Official Language & 3.9\% &
\ar{ههههههههه بيان مشترك قال ليك}
&
Hahahahaha, a ``joint statement'' you say. \\
\end{longtable}

Our results in Table 4 show that the public were generally more distrusting toward both the actual and perceived deepfakes, with lower trust and higher uncertainty in the perceived deepfake (Ardol's tweet) compared to the actual deepfake (Jonny's tweet). Contrary to \hyperlink{cite:vaccari2020}{Vaccari \& Chadwick (2020)}, more distrust was cited than uncertainty\footnote{We attribute this to the difficulty of operationalizing uncertainty in comments.}. The qualitative themes highlight that while people generally distrust AI deepfake tweets, there are many cited reasons for distrust; AI detection or the authenticity of the media is only one of them, and it is more recurring in the case of perceived/alleged deepfakes than in actual deepfakes. Prominent factors include source's reputation and the alignment of the narrative accompanying the footage with the user's political views.

\subsection{Perceived AI deepfake Case Study: Hmedti's Death}

The most prominent instance of AI skepticism in the case of Sudan is Hmedti's\footnote{Leader of Rapid Support Forces (RSF).} death propaganda\footnote{The impact of this propaganda was cumulative, making it difficult to identify a single most impactful tweet for analysis as the public views shifted over time.}. During the early days of the conflict, Sudanese Army Forces (SAF) supporters campaigned that Hmedti was dead, claiming that all evidence of his life was fake and AI-generated, leading the public to believe in his death, despite deepfake detection rejecting this claim (\hyperlink{cite:suliman2024}{Suliman, 2024}; [2]). While there have been many narratives (e.g., alternative character, Hmedti's twin), humanoid robots and AI-generated footage were among the notable ones. This incident has been influential on the public perception of AI deepfakery, especially since notable politicians have confirmed it. In their investigation of the psychological impact of Hmedti's death disinformation, \hyperlink{cite:beam2025b}{Beam Reports (2025b)} found that 33.1\% (n=557) felt happy after hearing Hmedti's death news. They also found that among those who have seen a video of him afterwards (n=378), 52.6\% believed the video was real, and the remaining 47.4\% either distrusted the video or were uncertain about it. Whether the propaganda's circulation is attributed to the supply-side motives or the demand-side vulnerabilities, namely the public's motivated reasoning and limited AI literacy, the results highlight the trend of using AI to reject dislikable truth [1]. This implies that in the case of Sudan, the epistemic threat is partially driven by perceived deepfakes.

\section{Reframing the AI Deepfake Problem}

\begin{longtable}{@{}>{\bfseries\RaggedRight}p{0.19\textwidth}@{\hspace{1em}}>{\RaggedRight}p{0.37\textwidth}@{\hspace{1em}}>{\RaggedRight\arraybackslash}p{0.37\textwidth}@{}}
\caption{Reframing the AI deepfake problem}\label{tab:reframing}\\
\toprule
\textbf{Angle} & \textbf{Current framing} & \textbf{What's missing} \\
\midrule
\endfirsthead
\multicolumn{3}{@{}l}{\textit{Table \thetable{} continued}}\\
\toprule
\textbf{Angle} & \textbf{Current framing} & \textbf{What's missing} \\
\midrule
\endhead
\bottomrule
\endfoot
AI deepfake risk &
\textbf{Actual AI deepfake}
\begin{itemize}[leftmargin=1em,itemsep=0pt,topsep=2pt]
\item Deepfakes promote belief in a falsified truth.
\item Malicious actors create high-quality deepfakes for manipulating the public.
\end{itemize}
$\rightarrow$ This fails to explain why the actual AI deepfake content in Sudan is relatively small, but still it poses a risk.
&
\textbf{Perceived AI deepfake}
\begin{itemize}[leftmargin=1em,itemsep=0pt,topsep=2pt]
\item Deepfake allegations can impact acceptance of objective truth.
\item People may refuse unfavorable, correct information by assuming that it might be deepfakes.
\end{itemize}
\\[6pt]
Disinformation Actor &
\textbf{Disinformed By Others}
\begin{itemize}[leftmargin=1em,itemsep=0pt,topsep=2pt]
\item The passive aggregator view suggests that the public passively absorbs disinformation from malicious actors.
\end{itemize}
$\rightarrow$ This fails to explain how the supply of AI deepfakes in Sudan is relatively low, but still people are AI skeptics.
&
\textbf{Disinformed by themselves}
\begin{itemize}[leftmargin=1em,itemsep=0pt,topsep=2pt]
\item The motivated public view argues that, instead of disinformation shaping beliefs, disinformation is shaped by existing demand side vulnerabilities such as motivated reasoning, cognitive effort and AI literacy.
\end{itemize}
\\
\end{longtable}

\subsection{Supply Side: Disinformed by Others}

The disinformed by others framing (supply side) argues that the problem stems from the production and the rising number of disinformation materials; applying it to the context of AI deepfakes, we can infer that the problem is attributed to the emerging use and volume of AI deepfakes in the Sudanese virtual propaganda. The supply side of AI deepfakes can be thought of in terms of the production of AI deepfakes to falsify truth (Jonny's tweet) and also the AI deepfake allegations to reject truth (Ardol's tweet). According to this viewpoint, those with corrupt motives mislead the public, who are supposedly objective in their media consumption. This view suggests interventions to suppress and control the production of AI deepfakes through laws and content moderation (\hyperlink{cite:schiffrin2017}{Schiffrin, 2017}; \hyperlink{cite:schiffrin2022}{Schiffrin, 2022}). This view fails to capture the AI deepfake impacts in Sudan given that the number of actual deepfake has been relatively small and the impact is amplified by perceived AI deepfakes.

\subsection{Demand Side: Misinformed by Themselves}

The motivated-by-themselves (demand side) of disinformation suggests that the problem is not attributed to the rising number of disinformation material, as they have always been there, but the problem has to do with consumption patterns that make users more vulnerable to it than ever (\hyperlink{cite:schiffrin2017}{Schiffrin, 2017}). Extending this to the context of deepfakes, we argue that the consumption patterns of disinformation in general are more concerning than the novelty or the adoption scale of the AI deepfake technology. This is evident by the limited number of AI deepfakes as compared to traditional disinformation in the Sudanese feeds [1]. In our analysis we treat the AI deepfake demand side as being influenced by motivated reasoning, cognitive efforts, and AI literacy, with motivated reasoning especially influential.

\subsubsection{Motivated Reasoning}

\begin{figure}[htbp]
\centering
\includegraphics[width=0.85\linewidth]{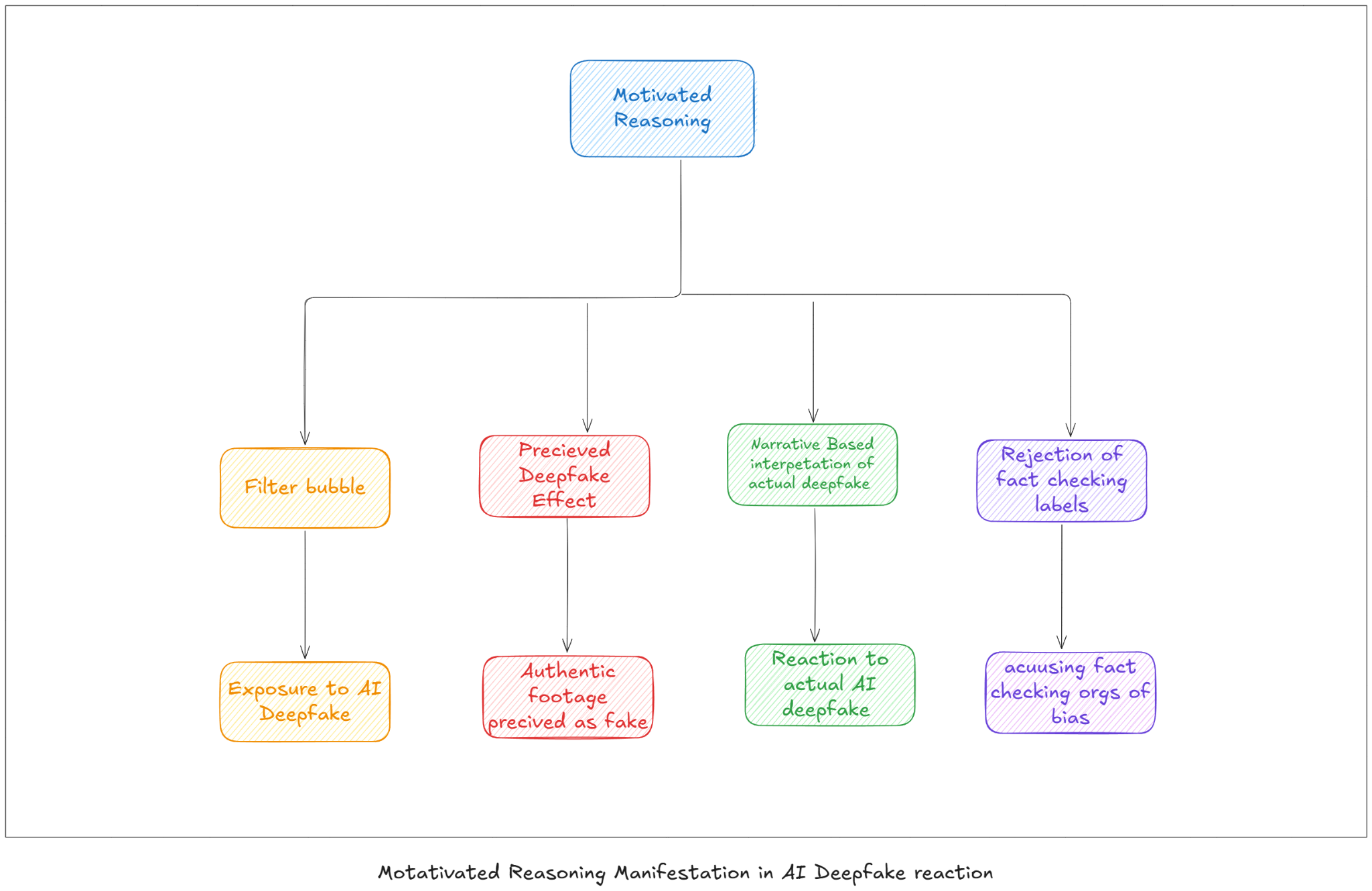}
\caption{Conceptual framework of motivated reasoning manifestation in reaction to AI deepfake}
\end{figure}

In our motivated reasoning framework, we outline four manifestations of motivated reasoning in AI deepfake reactions:

\begin{enumerate}[leftmargin=2em]

\item \textbf{Filter bubbles impact exposure to deepfake content.}

The first manifestation is important when looking at deepfakes in the context of social media algorithms, particularly the echo-chamber effect\footnote{This poses a limitation for our sampling. We discuss the limitations in more detail in the limitation section.}, as attention-seeking algorithms expose viewers to identity-confirming content\footnote{Improving the transparency of the algorithms would require collaboration from social media platforms, which we currently lack.}, which means their biases not only shape their truth judgement but also what they are exposed to in the first place.

\item \textbf{Motivated reasoning leading to the perception of authentic footage as deepfake (perceived deepfake).}

The second manifestation deals with what we refer to as a ``perceived AI deepfake.'' Ardol's tweet represents a textbook case of the liar dividend, where the deepfake allegation was made to protect the politician's reputation and thus relates to the liar dividend and the disinformed by others (supply) side, where the politician actively deceive the public to protect their reputation. This differs from Hmedti's death narrative, which was viral not only because of SAF supporters'\footnote{It is also often assumed to be Islamist-led propaganda.} led propaganda but also because segments of the public were inclined to believe it themselves, given their emotional attachment to the context. In other words, to some of them Hemdti's death would represent the punishment he deserves for what he has put the public through. For others his death may rearrange the power dynamics, wishing that it may end the war. All of this makes the motivated reasoning effect clearer in supplying and mobilizing deepfakes and blurs the line between ``others'' and themselves in causing the disinformation. As the public not only believed the narratives for themselves but also mobilized it and shared it with others. It's worth clarifying that the public may or may not be aware of the fact that they are deceiving themselves and others.

\item \textbf{Motivated reasoning and narratives impact reaction and interpretation of AI deepfake material.}

Motivated reasoning can also help us understand the reaction to actual and perceived deepfake content.\footnote{Crowd wisdom also impacts public reaction, as their views might conform to majority opinion.} The analysis of comments on Jonny's tweet, a technical AI deepfake footage, revealed a significant theme: motivated reasoning plays a huge role in tweet consumption as compared to footage truthfulness\footnote{This is known as confirmation bias, where people judge claims as more credible when those claims align with their prior beliefs or identities.} as seen by the ``agree with the media, disagree with the narrative'' theme. Comments on Ardol's tweet also showed how motivated reasoning shaped by previous beliefs---the actor's backing and reputation---might influence reaction towards the alleged deepfake.\footnote{The ``critical analysis/fact inconsistency'' might also relate to motivated reasoning. Refer to the limitations section.} This suggests that while the public mostly appears to be unaffected by the AI-generated deepfakes, this is a misleading sense of safety. In many cases, opposition to AI deepfake deception results from a lack of political alignment rather than AI detection skills. As a result, the public may be vulnerable to deception if future malicious actors deliberately create a realistic AI deepfake scenario and strategic publication campaigns.

\item \textbf{Rejecting AI deepfake correction and accusing fact checking organizations of being biased in their labelling}

The motivated reasoning of the public can constrain fact-check interventions to combat AI deepfakes when they are perceived to be biased in low-trust media contexts. Bias accusations of fact-checkers are already prevalent in the Sudanese media [1, 2]. This perception may stem from the longstanding lack of trust in the Sudanese media, which has long lacked objectivity. Alternatively, it may reflect the motivated reasoning of users who perceive fact checks to be biased when they do not align with their political views [1]. It's important to note that it's possible that human fact-checkers and individual civic tech practitioners become biased in some instances (\hyperlink{cite:johansson2023}{Johansson et al., 2023}; [2]) and in this case their motivated reasoning can be analyzed in the second and third levels of the framework.

\end{enumerate}

\subsubsection{Cognitive Effort}

While acknowledging limitations in our examination of cognitive efforts, we consider its influence well-supported by prior research. This view suggests that the public are lazy but objective in their consumption of deepfake content; therefore, interventions equip the public with the right information and encourage them to spend more cognitive effort in consuming content. While the analysis of comments in Jonny's tweet challenge this and reveal that the public are actually spending cognitive effort to align content with their belief, we don't think this is enough evidence to reject the impact of cognitive effort and we believe that the cognitive effort interventions accompanied by motivated reasoning interventions would be helpful for a wide range of users.

\subsubsection{AI Literacy}

While disinformation literature provides useful generalizable interventions, it is important to acknowledge the special context of AI deepfakes, as the visual and auditory elements may impact perceptions. This makes deepfake different from the other forms of disinformation. AI detection skills among the public in Sudan are assumed to be limited due to limited exposure and digital access, which may, in turn, lead to overestimation of the AI content generation capabilities [2].\footnote{ It's also possible that overestimation patterns continue even if some models have the capability due to the gap between these capabilities and AI accessibility and deployment in the context of Sudan.} This should be considered when thinking about AI skepticism. In this context, the visual/auditory evidence is discounted under the assumption that truth can be perfectly simulated by AI or at least assumed to be. However, while this is a plausible theory, our analysis of Ardol tweets showed contradictory pattern, as the majority rejected the liar's dividend claims\footnote{This might be in part due to lack of endorsement rather than genuine AI literacy.}. Whether the limited AI exposure indicates overestimation of AI capabilities or not, it relates to the evidence on the impact of AI self-efficacy in AI deepfake perception (\hyperlink{cite:liu2025}{Liu et al., 2025}) discussed earlier. Accordingly, integrating AI literacy should be prioritized in media literacy programs to build user capacity to detect AI-generated content. Moreover, there is a need to expand the number of AI detection tools that perform well on low resource languages and ensure tools' accessibility for local fact-checking organizations.

\section{Designing Interventions for AI Deepfakes}

It is important to acknowledge that AI deepfakes, while part of a broader disinformation ecosystem---one still dominated by prevailing cheap fakes and out-of-context fakes---pose acute risks due to their persuasive power. This suggests the need to design modality-agnostic interventions that generalize across different forms of disinformation. Simultaneously, AI-specific interventions, such as digital content transparency, still have gaps in their large-scale technical implementation, both in terms of our ``know-how'' and security flaws that can be used by malicious actors (\hyperlink{cite:nist2024}{NIST, 2024}; \hyperlink{cite:johansson2023}{Johansson et al., 2023}). These limitations suggest that technical interventions are not feasible in the short term to address AI deepfakes in the Sudanese media. In our investigation we found that the Sudanese efforts on combating AI deepfakes mostly rely on fact-checking by fact checking organizations or independent tech activists.

In this section we present interventions that are mainly borrowed from global disinformation literature. These modality agnostic interventions are adapted to the emerging dimensions of AI deepfakes and contextualised for Sudan's information environment. The rationale behind advocating for modality agnostic interventions is the limited number of AI deepfakes in Sudanese media. Therefore, instead of creating AI specific measures, it's more pragmatic to extend existing measures of disinformation.

\begin{longtable}{@{}>{\bfseries\RaggedRight}p{0.16\textwidth}@{\hspace{1em}}>{\RaggedRight}p{0.26\textwidth}@{\hspace{1em}}>{\RaggedRight}p{0.26\textwidth}@{\hspace{1em}}>{\RaggedRight\arraybackslash}p{0.26\textwidth}@{}}
\caption{Summary of Policy options}\label{tab:policy-options}\\
\toprule
& \textbf{Technical Interventions} & \textbf{Supply side (Regulation)} & \textbf{Demand Side (User empowerment)} \\
\midrule
\endfirsthead
\multicolumn{4}{@{}l}{\textit{Table \thetable{} continued}}\\
\toprule
& \textbf{Technical Interventions} & \textbf{Supply side (Regulation)} & \textbf{Demand Side (User empowerment)} \\
\midrule
\endhead
\bottomrule
\endfoot
\textbf{Focus} & Technical methods of documenting and accessing information about the origins and history of digital content. & Laws and content moderation to suppress and control the production of AI deepfakes. & Empower users with facts, AI literacy, and critical thinking skills. \\[4pt]
\textbf{Actor} & Research institutions & Governments and social media platforms & Fact-checking organizations, news agencies and tech activists \\[4pt]
\textbf{Prerequisites} & Technical know-how & Political will & Financial, human resources and access to users \\[4pt]
\textbf{Feasibility} & low & low & Relatively high \\
\end{longtable}

\subsection{Supply Side Interventions}

Supply side interventions focus on suppressing the production of deepfake by laws and content moderation policies. In Sudan, Article 66 of the Criminal Act of 1991 establishes the basis for mis/disinformation by criminalizing the spread of false information. Article 7 in the cybercrime law builds on this and criminalizes online disinformation with intent to ``cause fear or panic to the public,'' or ``threaten public peace or tranquility,'' or ``diminish the prestige of the state'' with four years of prison or flogging, or both (\hyperlink{cite:knight2023}{Knight \& Alsedeg, 2023}). These laws, while having the potential to address AI deepfakes, require amendments to establish clear enforcement mechanisms. There is also the gap between the written laws and the capacity and political will for their enforcement.

While social media platforms\footnote{Platform efforts to combat disinformation often collide with commitments towards freedom of expression (\hyperlink{cite:johansson2023}{Johansson et al., 2023}).} have made several attempts at content moderation, which can either lead to labeling, algorithmic downranking, delisting, removal, or deplatforming to mitigate disinformation, it is noted that these platforms spend a small magnitude of their safety budgets in global south regions (\hyperlink{cite:helmus2022}{Helmus, 2022a}; [2]). The disparity of safety policy among social media platforms makes it hard to scale and unify supply-side efforts across different platforms, particularly as X platform has been seen to be less cooperative with Sudanese fact-checkers than Meta and YouTube [1, 2]. Moreover, these platforms lack a comprehensive understanding of the Sudanese timeline and its nuances, which poses a barrier to their independent content moderation [1].

These supply-side interventions, however, require buy-in from governments and social media platforms, which we currently lack. The limited feasibility of the supply-side interventions, and the demand-led nature of AI deepfake risk in Sudan, motivate us to prioritize demand-side interventions.

\subsection{Demand Side Interventions}

These interventions operate on societal adaptation level, in other words it focuses on empowering users and building their resilience against produced deepfakes. We classify interventions that target demand side vulnerability based on the time of intervention as pre-event and post-event measures.

\subsubsection{Pre Event Measures}

\textbf{Pre-bunking.} It involves pre-event measures that expose users (demand side) to doses of disinformation in order for them to develop immunity against different manipulation techniques (e.g., inoculation games). These interventions have been shown to generalize across different political ideologies (\hyperlink{cite:roozenbeek2019}{Roozenbeek \& Van Der Linden, 2019}; \hyperlink{cite:roozenbeek2022}{Roozenbeek et al., 2022}). An AI deepfake-specific prebunking would include exposing the users to AI-generated content to develop detection skills and get a sense of the AI capabilities while also moderating the influence of political identity. Criticism of inoculation games includes that people were more skeptical and uncertain in general than they were able to distinguish truth from disinformation (\hyperlink{cite:johansson2023}{Johansson et al., 2023}). This calls for their conscious application in the context of Sudan, as it might lead to more AI skepticism.

\textbf{Media literacy.} A notable pre-event measure is media literacy, which involves educating users about disinformation and equipping them with skills to help them navigate digital spaces. It has proven effective, particularly in the context of breaking news and evaluating partisan views (\hyperlink{cite:johansson2023}{Johansson et al., 2023}). Despite concerns over its limited scalability, voluntary adoption, and risk of overconfidence in future truth evaluation (\hyperlink{cite:johansson2023}{Johansson et al., 2023}; [2]), media literacy remains one of the prominent proactive measures that empowers users to detect disinformation. In the context of AI deepfake mitigation, context-sensitive media literacy programs hold the promise of user empowerment, given that specific AI knowledge and toolkits have been integrated into these programs.

\subsubsection{Post Event Interventions}

\textbf{Fact-checking.} The most prominent post-event measure is fact-checking. Although fact-checking is reactive, post-event, and message-based, it has proven to be effective (\hyperlink{cite:ahmed2021}{Ahmed, 2021}; \hyperlink{cite:johansson2023}{Johansson et al., 2023}). In the context of Sudan, the only organization that is part of Meta's third-party fact-check program is Beam Reports\footnote{Beam Reports is a Sudanese fact-checking organization that focuses on empowering the public with fact-checking resources and educational materials to equip them with the necessary skills to navigate the complex media space in Sudan [1].} (\hyperlink{cite:suliman2024}{Suliman, 2024}). The program ensures that their fact-checking efforts translate into ``False'' labels and correction context on the platform [1].

Despite the potential of fact-checking efforts in combating AI deepfakes in Sudan, they have limitations relative to their scalability, visibility, and effectiveness. The scale problem with fact-checking lies in its dependency on the assessment of human fact-checkers. The limited reach of the fact-checking reports compared to the deepfake content poses the most important barrier for Beam Reports and other Sudanese fact-checking organizations to reach their audience (\hyperlink{cite:staff2018}{Staff, 2018}; [1, 2]). It is important to consider the digital nature of fact-checking reports, which contributes to their limited accessibility. The real and perceived AI deepfakes spread widely in Sudanese households through both digital and non-digital means; therefore, addressing them through digital means alone ignores how the digital divide complicates the nature of the problem (\hyperlink{cite:oliver2021}{Oliver, 2021}). The lingering impact of deepfakes represents a serious problem, as the public might resist correction and trust the initial communication more than subsequent communication\footnote{In the literature, it is known as the ``continued influence effect.''} due to heuristics, bias, or lack of alternative explanation (\hyperlink{cite:kahan2017}{Kahan, 2017a}; \hyperlink{cite:vanbavel2018}{Van Bavel \& Pereira, 2018}). This is further exacerbated by the visual nature of AI deepfakes. A problem Beam Reports found to be prevailing in their work in combating AI deepfakes is the challenge of visual memory correction [1].

\textbf{Debunking.} Debunking is a post-event intervention that clarifies the falsity of deepfakes by addressing the question of ``why they are fake,'' distinguishing it from fact-checking labels that merely provide a label. Research suggests that debunking is effective because it provides an alternative account in place of the deepfake (\hyperlink{cite:johansson2023}{Johansson et al., 2023}). The effectiveness of debunking depends on its reach and the trust placed in the actor responsible for the debunking. Moreover, there is research contesting the effectiveness of debunking, claiming that it has been shown to be impactful only in cases where participants held low or medium belief but not strong belief in the news (\hyperlink{cite:johansson2023}{Johansson et al., 2023}). This is relevant to the context of deepfakes in Sudan, as interventions must address both believers and skeptics of AI deepfakes. To ensure successful debunking, it is important to address the limited accessibility and reliability of official sources of information, a crucial challenge for Sudanese journalists and fact-checkers [1].

\textbf{Tiplines and self-help resources.} Closed-source environments\footnote{Closed-source environments are platforms where data is not publicly accessible (e.g., WhatsApp, Signal). Their limited accessibility represents a challenge for fact-checking efforts in these environments.} represent a critical dimension to consider when designing AI deepfake interventions. In Sudan, WhatsApp groups are considered incubators of disinformation and AI deepfake propaganda [1]. While it is technically possible to track deepfakes and disinformation in a closed-source environment, there seems to be limited acceptability for its practical application. This makes tiplines\footnote{Tiplines are accounts to which users can forward messages that they suspect to include false content, and as a result, they will be connected with fact-checking tools and organizations.} and self-help resources in WhatsApp very relevant for empowering users in these closed environments. An inspiring initiative was led by WhatsApp in partnership with Brazil's Election Authority (BEA), in which they launched a bot to which users could send messages that they suspected to be fake and receive fact-checks (\hyperlink{cite:johansson2023}{Johansson et al., 2023}). Transferring this initiative to the context of Sudan through partnerships with local fact-checking organizations will ensure public empowerment with fact-checking knowledge, even in closed-source environments. While tiplines can enable accessibility to existing fact-check databases, they lack scalability beyond the existing fact-check materials.

\textbf{AI-Assisted Interventions.} There have also been attempts to integrate AI in scaling the fact-checks, which results in enriching the self-help resources. AI tools with their wide accessibility decentralize interventions and amplify user empowerment\footnote{This happens in parallel with democratizing tools for producing AI deepfakes.}. Notably, there have been some attempts by the public in Sudan to use AI tools to detect AI-generated content or at least understand their current capabilities. In the comment analysis of Ardol's tweet, we found\footnote{Refer to the quote in the AI usage category.} that people used AI to gain a benchmark of AI model capability in generating content. The same pattern was seen by Mohamed Suliman---a tech activist---in one of the comments on one of his tweets. One in which he questioned the authenticity of an image that depicted a plane landing in Nyala's airport, and one of his followers asked Grok (X AI bot) to detect AI usage [2]. While research suggests that current Large Language Models (LLMs) are still not ready to detect deepfakes (\hyperlink{cite:tariq2025}{Tariq et al., 2025}), AI can provide new opportunities for tiplines and self-help interventions. In fact, efforts have already been made to train fact-checking AI models based on fact-check data on textual news with relatively high accuracy. Introducing AI detection and vision capabilities to AI fact-checking bots or current LLMs will represent significant progress in scaling fact-checking efforts and redirecting AI towards serving the truth.

\textbf{Reputational Ranking.} This mechanism is based on transaction cost economics to govern the exchange of information between a network of actors. It allows social media platforms to coordinate the exchange of information while allowing users to safeguard the information environment and promote accountability and trust (\hyperlink{cite:eccles2021}{Eccles \& Dingler, 2021}). Given the evidence from experimental research on the effectiveness of the politically balanced crowd trust ranking on distinguishing mainstream from hyperpartisan fake news sources across the political spectrum (\hyperlink{cite:pennycook2019}{Pennycook \& Rand, 2019b}), we argue that reputational ranking may reduce the influence of filter bubbles on the AI deepfake exposure. Reputational ranking also encourages reflective processing (\hyperlink{cite:johansson2023}{Johansson et al., 2023}), which in turn makes users alert to the influence of biases in their perception of content. The producers and mobilizers of deepfakes will be careful with their posts, as their reputation will be at stake, taking into account the active reaction to false labels by the authors of these posts [1]. The potential of reputational ranking stems from the responsibility associated with the act of ranking based on specific metrics, reputation, and truthfulness, as users will supposedly be very intentional in their rating and, most importantly, their sharing, which distinguishes it from the current model of engagement-based algorithmic amplification\footnote{Platforms rank posts based on engagement (like/share/comments).}. However the cost of user experience\footnote{If truthfulness ratings are mandatory, user dropout could rise---discouraging platform participation; if ratings are voluntary, meaningful uptake will likely require incentives to motivate sustained effort.} is important to consider for accurate analysis of the feasibility and effectiveness of this intervention.

\section{Recommendations}

\begin{enumerate}[leftmargin=2em]
\item Empower fact-checking organizations with resources, particularly:
\begin{enumerate}[leftmargin=2em]
\item \textbf{Technical resources:} AI detection tools that comprehend the Sudanese dialect and other fact-checking tools, will empower them with the technical competencies to navigate the digital space.
\item \textbf{Reach:} Increased visibility and accessibility of their fact-checking pieces not only through digital means but also through non-digital means such as TV channels, radio outlets, and community outreach will ensure that AI deepfake reports reach different societal components (\hyperlink{cite:oliver2021}{Oliver, 2021}).
\item \textbf{Sustainable funds:} Direct long-term philanthropic and diverse base funds for fact-checking organizations will help ensure their independence and sufficiency (\hyperlink{cite:oliver2021}{Oliver, 2021}).
\end{enumerate}
These steps are crucial for scaling the impact of existing organizationslike BeamReports and Jonuina, but most importantly, for expanding the fact-checking ecosystem in Sudan by encouraging new actors to join.

\item Provide resources for wide-scale media literacy campaigns on AI deepfakes, their potential usage, and toolkits to mitigate their impact. These efforts should target a broad audience through innovative ideas that combine education with accessibility [1, 2]. To ensure that both cognitive and motivated reasoning vulnerabilities are addressed, these media literacy campaigns should be designed to uplevel critical thinking, reflexivity anf AI detection skills.

\item Introduce tiplines and self-help resources (e.g., Brazil bot) in closed-source environments through strategic collaboration between WhatsApp and local fact-checking organizations that ensures scalability and ease of access to fact-checking reports.

\item Integrate AI in scaling deepfake detection and media literacy. More research attention should be given to the adoption of AI to design wide-scale fact checking interventions. This will require collaboration from social media platforms, fact-checking organizations, and AI research labs, each bringing specific expertise to turn AI into a force for truth.

\item Broader advocacy---including academic institutions, multilateral bodies, media agencies, and civil society---should call for more algorithmic transparency and seek the platform's commitment to implementing reputational ranking. Such long-term structural reform helps realign platform incentives for accuracy and accountability. Advocacy efforts should also call for increased safety spending by social media platforms in fragile states.
\end{enumerate}

\section{Limitations}

\textbf{Measurement via comments.} Using comments as a proxy for truth judgement is the critical limitation in our investigation. Comments are not specific answers to the prompt ``Do you trust the footage, and why?'' Therefore, as event-centered methodologies suggest, comments might reflect partial and irrelevant views. Relating the comments to the themes and truth judgement categories required interpretive coding. To ensure reliability for our coding, we performed three coding passes and documented assumptions in the Annex C. Future research might consider developing precise tunnel-based measures for the trust in the AI deepfake disinformation. Moreover, experiments that moderate the influence of individual-level factors (motivated reasoning and cognitive abilities) would be valuable in advancing our understanding of the public reaction.

\textbf{Coding Subjectivity.} In labelling ``Critical analysis/fact inconsistency,'' it is important to acknowledge that the perception of fact is subjective, or at least it depends on the comment tone. This is an area where the researcher bias might also be relevant. We tried to mitigate this by providing a detailed coding framework in Annex C.

\textbf{Platform scope.} Using an open-source environment, particularly X, poses a barrier for the external validity of our findings. Future research may consider working on cross-platform analysis to compare the impact of platform norms and user reactions across platforms.

\textbf{Sampling and algorithmic exposure.} The nature of social media algorithms poses a barrier for our sampling, as the exposure to the AI deepfake tweets might already be influenced by existing biases. Moreover those who place comments may have different characteristics (e.g. more political or vocal) than the users who viewed the tweets without placing comments. Therefore the motivated reasoning influence may not be as high as the reported influence for non commenters. Future research might consider employing random sampling for users and manipulating their exposure to AI deepfake content through controlled trials. That said, the generated synthetic media should be controlled so that it is not propagated elsewhere in different contexts.

\textbf{Individual Factors \& Networks Excluded.} Due to limited resources, the research does not analyze the impact of group-level factors and individual-level factors (e.g., popularity, political agenda, etc.) on deepfake perception. Future research should consider the impact of these factors and intergroup relationships to examine the influence of profiles and networks on the reaction and exposure.

\section *{Note on AI usage}
General-purpose AI models have been used for coding, editing and formatting the manuscript.
\section*{References}

\refentry{ahmed2021}{Ahmed, S. (2021). Fooled by the fakes: Cognitive differences in perceived claim accuracy and sharing intention of non-political deepfakes. \textit{Personality and Individual Differences}, \textit{182}, 111074. \url{https://doi.org/10.1016/j.paid.2021.111074}}

\refentry{beam2025a}{Beam Reports. (2025a, May 19). Artificial Intelligence: A New Chapter in the War of Misinformation in the Sudanese Digital Space. \textit{Beam Reports}. Retrieved August 7, 2025, from \url{https://en.beamreports.com/21374/}}

\refentry{beam2025b}{Beam Reports. (2025b, February 24). \textit{Al-Athar al-Nafsi li-l-Ma'allimāt al-Mudallalah: ``Mawt Ḥumaydī'' Namūdhajan} [The psychological impact of disinformation: The ``Death of Ḥumaydī'' as a case study]. Sudalytica. \url{https://sudalytica.beamreports.com/}}

\refentry{byman2023}{Byman, D. L., Gao, C., Meserole, C., Subrahmanian, V. S., \& Foreign Policy at Brookings. (2023). \textit{Deepfakes and international conflict}. \url{https://www.brookings.edu/wp-content/uploads/2023/01/FP_20230105_deepfakes_international_conflict.pdf}}

\refentry{eccles2021}{Eccles, D. A., \& Dingler, T. (2021b). Three prophylactic interventions to counter fake news on social media. \textit{arXiv (Cornell University)}. \url{https://doi.org/10.48550/arxiv.2105.08929}}

\refentry{gambin2024}{Gambín, Á. F., Yazidi, A., Vasilakos, A. et al. (2024). Deepfakes: current and future trends. \textit{Artificial Intelligence Review}, \textit{57}, 64. \url{https://doi.org/10.1007/s10462-023-10679-x}}

\refentry{goodman2023}{Goodman, J., \& Hashim, M. (2023, October 5). \textit{AI: Voice cloning tech emerges in Sudan civil war}. BBC News. Retrieved August 7, 2025, from \url{https://www.bbc.com/news/world-africa-66987869}}

\refentry{helmus2022}{Helmus, T. C. (2022a). Artificial Intelligence, deepfakes, and disinformation: A primer. In \textit{RAND Corporation eBooks}. \url{https://doi.org/10.7249/pea1043-1}}

\refentry{johansson2023}{Johansson, P., Enoch, F., Hale, S. A., Vidgen, B., Bereskin, C., Margetts, H. Z., \& Bright, J. (2023, November 1). How can we combat online misinformation? A systematic overview of current interventions and their efficacy. Available at SSRN: \url{https://ssrn.com/abstract=4648332} or \url{http://dx.doi.org/10.2139/ssrn.4648332}}

\refentry{roozenbeek2022}{Roozenbeek, J., et al. (2022). Psychological inoculation improves resilience against misinformation on social media. \textit{Science Advances}, \textit{8}, eabo6254. \url{https://doi.org/10.1126/sciadv.abo6254}}

\refentry{kahan2017}{Kahan, D. M. (2017). Misconceptions, misinformation, and the logic of identity-protective cognition. \textit{SSRN Electronic Journal}. \url{https://doi.org/10.2139/ssrn.2973067}}

\refentry{khalafallah2025}{Khalafallah, H. (2025, January 14). Beyond the battlefield: Sudan's virtual propaganda warzone. \textit{The Tahrir Institute for Middle East Policy}. Retrieved August 7, 2025, from \url{https://timep.org/2025/01/14/beyond-the-battlefield-sudans-virtual-propaganda-warzone/}}

\refentry{knight2023}{Knight, T., \& Alsedeg, L. (2023). \textit{Democracy derailed: Sudan's precarious information environment, 2019--2022} (I. Robertson \& A. Carvin, Eds.). Atlantic Council. \url{http://www.jstor.org/stable/resrep52679}}

\refentry{liu2025}{Liu, Q., Wang, L., \& Luo, M. (2025). When seeing is not believing: self-efficacy and cynicism in the era of intelligent media. \textit{Humanities and Social Sciences Communications}, \textit{12}, 274. \url{https://doi.org/10.1057/s41599-025-04594-5}}

\refentry{nist2024}{NIST, G. M. (2024). Reducing risks posed by synthetic content, An overview of technical approaches to digital content transparency. \url{https://doi.org/10.6028/nist.ai.100-4}}

\refentry{oliver2021}{Oliver, L. (2021, September 30). \textit{The fight for facts in the Global South: How four projects are building a new model}. Reuters Institute for the Study of Journalism. \url{https://reutersinstitute.politics.ox.ac.uk/news/fight-facts-global-south-how-four-projects-are-building-new-model}}

\refentry{pennycook2018}{Pennycook, G., \& Rand, D. G. (2018). Lazy, not biased: Susceptibility to partisan fake news is better explained by lack of reasoning than by motivated reasoning. \textit{Cognition}, \textit{188}, 39--50. \url{https://doi.org/10.1016/j.cognition.2018.06.011}}

\refentry{pennycook2019}{Pennycook, G., \& Rand, D. G. (2019). Fighting misinformation on social media using crowdsourced judgments of news source quality. \textit{Proceedings of the National Academy of Sciences}, \textit{116}(7), 2521--2526. \url{https://doi.org/10.1073/pnas.1806781116}}

\refentry{roozenbeek2019}{Roozenbeek, J., \& van der Linden, S. (2019). Fake news game confers psychological resistance against online misinformation. \textit{Palgrave Communications}, \textit{5}, 65. \url{https://doi.org/10.1057/s41599-019-0279-9}}

\refentry{schiffrin2022}{Schiffrin, A. (2022). The pursuit of truth: Fixes for the spread of online mis/disinformation. In \textit{Columbia SIPA Institute of Global Politics}. \url{https://igp.sipa.columbia.edu/sites/igp/files/2023-12/IGP_Anya_Schiffrin_The_Pursuit_of_Truth-Fixes_for_the_Spread_of_Online_Mis_Disinformation.pdf}}

\refentry{schiffrin2017}{Schiffrin, A. (2017, October 26). \textit{How Europe fights fake news}. Columbia Journalism Review. \url{https://www.cjr.org/watchdog/europe-fights-fake-news-facebook-twitter-google.php}}

\refentry{schiff2024}{Schiff, K. J., Schiff, D. S., \& Bueno, N. S. (2024). The liar's dividend: Can politicians claim misinformation to evade accountability? \textit{American Political Science Review}, \textit{119}(1), 71--90. \url{https://doi.org/10.1017/S0003055423001454}}

\refentry{schwartz2018}{Schwartz, O. (2018, November 12). \textit{You thought fake news was bad? Deep fakes are where truth goes to die.} The Guardian. \url{https://www.theguardian.com/technology/2018/nov/12/deep-fakes-fake-news-truth}}

\refentry{suliman2024}{Suliman, M. (2024, October 23). \textit{The deepfake is a powerful weapon in the war in Sudan}. African Arguments. Retrieved August 7, 2025, from \url{https://africanarguments.org/2024/10/the-deepfake-is-a-powerful-weapon-in-the-war-in-sudan/}}

\refentry{staff2018}{Staff, C. (2018b, October 10). \textit{Issue brief: The `demand side' of the disinformation crisis}. National Endowment for Democracy. \url{https://www.ned.org/issue-brief-the-demand-side-of-the-disinformation-crisis/}}

\refentry{tariq2025}{Tariq, S., Nguyen, D., Chamikara, M. A. P., Wu, T., Abuadbba, A., \& Moore, K. (2025, June 12). \textit{LLMs are not yet ready for deepfake image detection}. arXiv.org. \url{https://arxiv.org/abs/2506.10474}}

\refentry{vaccari2020}{Vaccari, C., \& Chadwick, A. (2020). Deepfakes and disinformation: Exploring the impact of synthetic political video on deception, uncertainty, and trust in news. \textit{Social Media + Society}, \textit{6}(1). \url{https://doi.org/10.1177/2056305120903408}}

\refentry{vanbavel2018}{Van Bavel, J. J., \& Pereira, A. (2018, January). \textit{The partisan brain: An identity-based model of political belief} [Preprint]. PsyArXiv. \url{https://doi.org/10.31234/osf.io/ak642}}

\clearpage
\appendix

\section{Annex A: Interviews Citation}

\begin{longtable}{@{}>{\RaggedRight}p{0.05\textwidth}@{\hspace{1em}}>{\RaggedRight}p{0.38\textwidth}@{\hspace{1em}}>{\RaggedRight}p{0.20\textwidth}@{\hspace{1em}}>{\RaggedRight\arraybackslash}p{0.24\textwidth}@{}}
\caption{Interview citations}\label{tab:annexA}\\
\toprule
\textbf{No.} & \textbf{Interview Citation} & \textbf{Interviewee Name} & \textbf{Expertise/Role} \\
\midrule
\endfirsthead
\multicolumn{4}{@{}l}{\textit{Table \thetable{} continued}}\\
\toprule
\textbf{No.} & \textbf{Interview Citation} & \textbf{Interviewee Name} & \textbf{Expertise/Role} \\
\midrule
\endhead
\bottomrule
\endfoot
{[}1{]} & (Beam Reports, personal communication, August 26, 2025) & Beam Reports & Fact-checking Organization \\
{[}2{]} & (M. Suliman, personal communication, August 25, 2025) & Mohamed Suliman & Disinformation and AI safety Researcher \\
\end{longtable}

\section{Annex B: Tweet Citation}

\begin{longtable}{@{}>{\RaggedRight}p{0.14\textwidth}@{\hspace{1em}}>{\RaggedRight}p{0.53\textwidth}@{\hspace{1em}}>{\RaggedRight\arraybackslash}p{0.20\textwidth}@{}}
\caption{Tweet citations}\label{tab:annexB}\\
\toprule
\textbf{Tweet} & \textbf{Citation} & \textbf{Google Drive Link} \\
\midrule
\endfirsthead
\multicolumn{3}{@{}l}{\textit{Table \thetable{} continued}}\\
\toprule
\textbf{Tweet} & \textbf{Citation} & \textbf{Google Drive Link} \\
\midrule
\endhead
\bottomrule
\endfoot
Transitional council tweet & Transitional Sovereignty Council -- Sudan [@TSC\_SUDAN]. (2024, October 4). [Post]. X. \url{https://x.com/tsc_sudan/status/1842100027860156901} & \url{https://drive.google.com/file/d/1T3QIP_HzGgKQkPJ_ogOKF2tSvl6TbWR8/view?usp=sharing} \\[4pt]
Jonny's tweet & Gould, J. [@jonnygould]. (2024, March 14). EXCLUSIVE: Abdel Fattah al-Burhan, head of Muslim Brotherhood in Sudan is ordering his terrorists to attack their own civilians\ldots{} [Post]. X. \url{https://x.com/jonnygould/status/1768345232184148139} & \url{https://drive.google.com/drive/folders/1X9iypiB_K4fz8YXC-JazIbpJAnXvimHj?usp=sharing} \\[4pt]
Salah\_Mana3 real tweet & Manaa, S. [@SalahManaa5]. (2024a, March 9). \ar{مليشيات علي كرتي يقودها البرهان باسم الجيش الوطني وهي تحتفل بانها مليشيات داعشية ولا تؤمن بالدولة السودانية} [Post]. X. \url{https://x.com/SalahManaa5/status/1766326622087512156} & \url{https://drive.google.com/file/d/1TRPD_KNtMJ6swkXM8MOY9lvt40s3ulVM/view?usp=sharing} \\[4pt]
Salah\_Mana3 AI tweet & Manaa, S. [@SalahManaa5]. (2024b, March 15). \ar{ظهر الحق وزهق الباطل ان الباطل كان زهوقا البرهان قائد مليشيات علي كرتي يأمر بقتل المدنيين} [Post]. X. \url{https://x.com/SalahManaa5/status/1768530517496905886} & \url{https://drive.google.com/file/d/1rB3HRhtNcV_Zu6Gm9xo7XfsTGP4RwGBg/view?usp=sharing} \\[4pt]
Ardol tweet & Ardol, M. [@MubarakArdol]. (2023, July 17). \ar{بيان مشترك للرأي العام هنالك تسجيل متداول في السوشيال ميديا تبثه الغرف المعادية ومعمول بصورة فنية متقنة}\ldots{} [Post]. X. \url{https://x.com/MubarakArdol/status/1680602263352475650} & \url{https://drive.google.com/drive/folders/1JuUC7zeW7a7IHfoqIxZRRQsyNpdprHUm?usp=sharing} \\[4pt]
Mohamed Suliman tweet & Kambal, M. [@MuhammedKambal]. (2025, August 8). This image seems AI-generated, no? [Post]. X. \url{https://x.com/MuhammedKambal/status/1953854738744611207} & \url{https://drive.google.com/drive/folders/1fmaE6JHURg8xUYyfC1pFDJotbgdgzfoj?usp=sharing} \\
\end{longtable}

\section{Annex C: Coding Framework}

\textbf{Rules and assumptions for thematic labelling}

\textbf{Jonny's tweet}

\begin{enumerate}[leftmargin=2em]

\item For those who distrusted the narrative and either did not explicitly express distrust in the footage or expressed trust in the footage, the label ``Agree with the footage but disagree with the context/narrative'' was given to them. Other themes accompanied it to justify it. Note that this theme was used to label theme 1 only which is an important caveat when comparing its frequency with other themes that was given for either theme 1, 2 or 3.

\item The AI usage label was given to comments that explicitly referred to the fabricated nature of the footage and/or AI manipulation.

\item The ``Agree with the footage but disagree with the context/narrative'' theme appeared frequently with the ``critical analysis/fact inconsistency'' and ``support for the subject of the recording'' themes. In this setup:
\begin{enumerate}[leftmargin=2em]
\item The label ``Critical analysis/fact inconsistency'' was given to comments that either:
\begin{enumerate}[leftmargin=2em]
\item corrected the statement that al-Burhan is the head of the Islamic Brotherhood,
\item contextualized the specific operation/location discussed,
\item shared RSF violations and the incidents they have witnessed\footnote{Refer to the potential limitation of this labelling rule in the limitations section.} (e.g., ``The Sudanese Armed Forces are fighting the Rapid Support Forces that occupied the homes of citizens. As a Sudanese citizen, the Rapid Support Forces occupied my house.''),
\item comments on the inaccuracy of the translation.
\end{enumerate}
\item ``Support for the subject of the recording'' was given to:
\begin{enumerate}[leftmargin=2em]
\item comments that expressed explicit support for the actor,
\item comments expressing opposition to the actor's rivalry.
\end{enumerate}
\end{enumerate}

\item The ``source distrust'' label either regarded him as an external party with no knowledge of the context, paid by RSF or UAE, lacked fact checks, or viewed the tweet as part of Israeli-led propaganda.

\item Comments with hashtags like ``\#\ar{دبل\_ليهو}'' and ``\#\ar{بل\_بس}'', which were associated with SAF supporters, were preserved and labeled as ``Support for the subject of the recording.''

\end{enumerate}

\textbf{Ardol's tweet}

\begin{enumerate}[leftmargin=2em]
\item The ``distrusting'' label in this tweet refers to distrust in the tweet, which assumed AI usage. Put differently, the label captured trust in the authenticity of the media and distrust in the AI deepfake narrative.
\item Comments with references to statements in the video were labeled as ``distrusting,'' and their theme was ``Others.''
\item The ``critical analysis/fact inconsistency'' label was given to comments that disbelieved AI usage because the recording content matches prior facts.
\item Other labels were self-explanatory.
\end{enumerate}

\end{document}